\documentclass[submitting]{nst}

\usepackage{threeparttable}
\usepackage{subfigure,dcolumn}
\usepackage{epstopdf}
\usepackage{mhchem}
\usepackage{upgreek}
\usepackage{color}
\par
\begin{document}

\title{$\alpha$-cluster decay from $^{24}$Mg resonances produced in $^{12}$C($^{16}$O,$^{24}$Mg)$\alpha$ reaction}
\thanks{Supported by the National Key R\&D Program of China (2018YFA0404403), the National Natural Science Foundation of China (11875074, 11875073, 12027809, 11961141003, U1967201, 11775004, and 11775003), the Continuous Basic Scientific Research Project (WDJC-2019-13), and the State Key Laboratory of Nuclear Physics and Technology, Peking University (NPT2020KFY10).}

\author{Dong-Xi Wang}
\affiliation{School of Physics and State Key Laboratory of Nuclear Physics and Technology, Peking University, Beijing 100871, China}

\author{Yan-Lin Ye}
\email[Corresponding author, ]{yeyl@pku.edu.cn}
\affiliation{School of Physics and State Key Laboratory of Nuclear Physics and Technology, Peking University, Beijing 100871, China}

\author{Cheng-Jian Lin}
\affiliation{Department of Nuclear Physics, China Institute of Atomic Energy, Beijing 102413, China}

\author{Jia-Hao Chen}
\affiliation{School of Physics and State Key Laboratory of Nuclear Physics and Technology, Peking University, Beijing 100871, China}

\author{Kai Ma}
\affiliation{School of Physics and State Key Laboratory of Nuclear Physics and Technology, Peking University, Beijing 100871, China}

\author{Jia-Xing Han}
\affiliation{School of Physics and State Key Laboratory of Nuclear Physics and Technology, Peking University, Beijing 100871, China}

\author{Hui-Ming Jia}
\affiliation{Department of Nuclear Physics, China Institute of Atomic Energy, Beijing 102413, China}

\author{Lei Yang}
\affiliation{Department of Nuclear Physics, China Institute of Atomic Energy, Beijing 102413, China}

\author{Li-Sheng Yang}
\affiliation{School of Physics and State Key Laboratory of Nuclear Physics and Technology, Peking University, Beijing 100871, China}

\author{Zi-Yao Hu}
\affiliation{School of Physics and State Key Laboratory of Nuclear Physics and Technology, Peking University, Beijing 100871, China}

\author{Ying Chen}
\affiliation{School of Physics and State Key Laboratory of Nuclear Physics and Technology, Peking University, Beijing 100871, China}

\author{Wei-Liang Pu}
\affiliation{School of Physics and State Key Laboratory of Nuclear Physics and Technology, Peking University, Beijing 100871, China}

\author{Gen Li}
\affiliation{School of Physics and State Key Laboratory of Nuclear Physics and Technology, Peking University, Beijing 100871, China}

\author{Zhi-Wei Tan}
\affiliation{School of Physics and State Key Laboratory of Nuclear Physics and Technology, Peking University, Beijing 100871, China}

\author{Hong-Yu Zhu}
\affiliation{School of Physics and State Key Laboratory of Nuclear Physics and Technology, Peking University, Beijing 100871, China}

\author{Tian-Peng Luo}
\affiliation{Department of Nuclear Physics, China Institute of Atomic Energy, Beijing 102413, China}

\author{Shan-Hao Zhong}
\affiliation{Department of Nuclear Physics, China Institute of Atomic Energy, Beijing 102413, China}

\author{Da-Hu Huang}
\affiliation{Department of Nuclear Physics, China Institute of Atomic Energy, Beijing 102413, China}

\author{Jian-Ling Lou}
\affiliation{School of Physics and State Key Laboratory of Nuclear Physics and Technology, Peking University, Beijing 100871, China}

\author{Xiao-Fei Yang}
\affiliation{School of Physics and State Key Laboratory of Nuclear Physics and Technology, Peking University, Beijing 100871, China}

\author{Qi-Te Li}
\affiliation{School of Physics and State Key Laboratory of Nuclear Physics and Technology, Peking University, Beijing 100871, China}

\author{Jin-Yan Xu}
\affiliation{School of Physics and State Key Laboratory of Nuclear Physics and Technology, Peking University, Beijing 100871, China}

\author{Zai-Hong Yang}
\affiliation{School of Physics and State Key Laboratory of Nuclear Physics and Technology, Peking University, Beijing 100871, China}

\author{Kang Wei}
\affiliation{School of Physics and State Key Laboratory of Nuclear Physics and Technology, Peking University, Beijing 100871, China}

\begin{abstract}
A transfer reaction and cluster-decay experiment, $^{12}$C($^{16}$O,$^{24}$Mg$\rightarrow$$\alpha$+$^{20}$Ne)$\alpha$, was performed at a beam energy of 96 MeV. Both recoil and decay $\alpha$ particles were detected in coincidence, allowing us to deduce the energy-momentum of a $^{20}$Ne fragment. A number of resonant states of $^{24}$Mg were reconstructed up to an excitation energy of approximately 30 MeV. Owing to the experimentally achieved excellent resolutions of the $Q$-value and excitation-energy spectra, the relative decay widths for each resonant state in $^{24}$Mg to various final states of $^{20}$Ne were extracted, alone with the total decay width. The obtained results provide good testing ground for theoretical descriptions of the multiple clustering configurations in $^{24}$Mg.
\end{abstract}

\keywords{reaction Q-value, transfer reaction, cluster decay, relative decay width}

\maketitle

\section{INTRODUCTION}\label{sec.I}

Clustering is a fundamental phenomenon of the nuclear structure, which tends to emerge around the corresponding cluster-separation threshold and may persist to higher excitation energies~\cite{PTPSE68464, PR43243, PPNP8278, RMP90035004, NST291}. Originally, inclusive missing-mass measurements were used to speculate the possible cluster structure by excluding the states already classified into the single-particle systematics ~\cite{PR43243,NST291}. Over the past two decades, the coincident measurement of decay fragments has allowed the selection and reconstruction of resonant states that possess relatively large cluster probability, therefore avoiding the high level density of the single-particle-type states~\cite{NST291,SSPMA50112003}. To date, considerable theoretical and experimental research has been devoted to the study of cluster configurations in neutron-rich beryllium, carbon and oxygen isotopes~\cite{PRL110262501, PRL112162501, PRL113032506, SCPM571613, PRC91024304,CPM60062011,PRC95021303,CPC42074003, PRC99064315,CPC43084001,SCPM6212011,NST29184,PRL124192501,PRC101031304,PRC105044302,
CPC40111001}.

Moving to heavier nuclei, clustering in $^{24}$Mg is interesting because of not only the richness of its cluster configurations, but also its role in nuclear astrophysics associated with the formation of elements in stars, such as carbon burning~\cite{PRC76035802, CROLFs1988}. The cluster configurations of $^{24}$Mg include $\alpha$ + $^{20}$Ne, $^{12}$C + $^{12}$C, $^{16}$O + 2$\alpha$, $^{12}$C + 3$\alpha$ and the 6$\alpha$ condensation state, with increasing separation thresholds at 9.32, 13.93, 14.05, 21.21, and 28.48 MeV, respectively. These configurations have been investigated theoretically and experimentally but with limited robust outcomes, especially in the high excitation region ~\cite{SCPMA54130,PRC99014606,NPA463399, NPA475219, PLB2286, NPA709275, PRC91061302,PRC271550,PLB267325,PLB181233,PLB174246,PLB181299,PRC571277,PRC68054321,NPCCP257,PRC103044315,PRL129102701}.

One $\alpha$ emission ($\alpha$+$^{20}$Ne) corresponds to the lowest separation energy among all cluster-decay channels of $^{24}$Mg and hence should appear strongly in measurements. It is thus important for theoretical models to correctly describe the $\alpha$+$^{20}$Ne structure in $^{24}$Mg together with the associated decay paths and widths before being applied to other more complicated channels. So far, several experiments have identified a number of $\alpha$+$^{20}$Ne clustering states in $^{24}$Mg for excitation energies below 18.5 MeV, as summarized in Refs.\cite{PRC432523,PRC452693} and comparatively listed in the last column of Table \ref{tab1}.  These measurements were inclusive and generated significantly more states than those with strong clustering structures. To be more selective, it is necessary to detect and identify the decay $\alpha$ particle or $^{20}$Ne fragment. This was once experimentally realized and reported in several conferences~\cite{JPCS436012009, PRC103044315}, where a few $0^+$ states were tentatively allocated, as also listed in Table \ref{tab1}.

In this paper, we present a new measurement of the reaction-decay process $^{16}$O($^{12}$C, $^{24}$Mg$\rightarrow$$\alpha$+$^{20}$Ne)$\alpha$ for $^{24}$Mg excitation up to 30 MeV. The coincidentally detected data allow to reconstruct the clustering resonant states according to the decay paths related to various final states of the $^{20}$Ne fragment. The results may be important as a benchmark for the future theoretical calculations.

\section{DESCRIPTION OF THE EXPERIMENT}\label{sec.II}

\begin{figure}[!htb]
\includegraphics[width=7 cm]{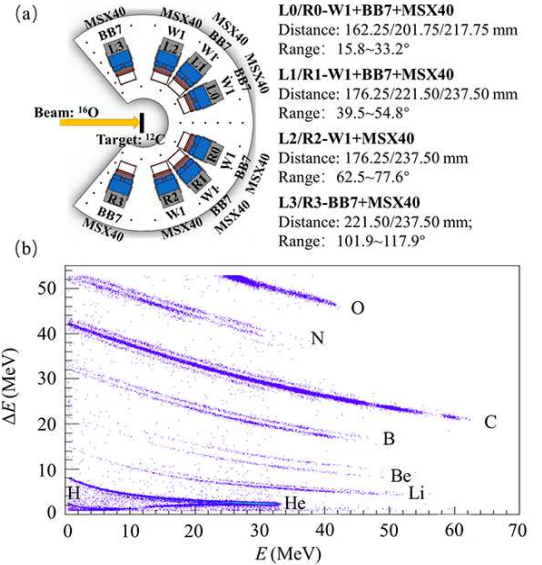}
\caption{(a) Schematic view of the experimental setup. The distances of the detector layers, such as W1+BB7+MSX40, for each telescope pair, such as L0/R0, are indicated alone with the corresponding angular coverage of the telescope. (b) Particle identification using the standard method of the energy loss ($\Delta E$) versus the remaining energy ($E$) taken from one pixel of the L0-DSSDs..}
\label{fig1}
\end{figure}

The experiment was performed at the HI-13 tandem accelerator at the China Institute of Atomic Energy (CIAE) in Beijing. A 96-MeV $^{16}$O beam with an intensity of of approximately 4.5 enA was incident on a 220 $\mu$g/cm$^{2}$ (0.98 $\mu$m) self-supporting $^{\rm nat.}$C target. The transfer reaction $^{12}$C($^{16}$O,$^{24}$Mg$^{*}$)$\alpha$ was used to populate the high lying states in $^{24}$Mg. Both recoil and decay $\alpha$ particles were detected by an array of eight position sensitive charged-particle telescopes, namely L0$\sim$3, and R0$\sim$3, which were symmetrically placed on both sides of the beam axis, as schematically displayed in Fig.~\ref{fig1}(a).

\begin{table*}
\centering
\caption{ \label{tab1}  Summary of the resonant states of $^{24}$Mg reconstructed from the $\alpha$+$^{20}$Ne decay channel. For comparison, the corresponding results from previous AMD calculations and inelastic or resonant scattering measurements are also presented.}
\footnotesize
\begin{tabular*}{170mm}{@{\extracolsep{\fill}}ccccccccccc}
\toprule
\multicolumn{5}{c}{this work} & \multicolumn{2}{c}{AMD~\cite{PRC91061302}} & \multicolumn{2}{c}{Inelastic exp.}& \multicolumn{2}{c}{Resonant exp.~\cite{PRC432523}}\\
\cmidrule(r){1-5} \cmidrule(r){6-7} \cmidrule(r){8-9} \cmidrule(r){10-11}
$E_{x}$/ & $\Gamma$/ & \multicolumn{3}{c}{Relative Decay Width}  &$E_{x}$/  &$J_{\pi}$ &$E_{x}$/    &$J_{\pi}$ &$E_{x}$/    &$J_{\pi}$   \\
MeV     & keV      &  $^{20}$Ne (g.s.)  &  $^{20}$Ne (2$^{+}$)  &  $^{20}$Ne (4$^{+}$)  &MeV        &          &MeV        &          &MeV        &            \\
\cmidrule(r){1-5} \cmidrule(r){6-7} \cmidrule(r){8-9} \cmidrule(r){10-11}
12.6 (1) 	&260 (20) 	&1.00  	&   	&   	&		&				&		&			&12.578 	&2$^{+}$		\\
13.1 (1) 	&140 (20) 	&0.56 	&0.44 	&   	&13.2 	&0$_{3}^{+}$	&13.1~\cite{JPCS436012009} 	&0$^{+}$	&13.089 	&2$^{+}$		\\
13.7 (1) 	&180 (30) 	&0.42 	&0.58 	&   	&		&				&13.79 (1)~\cite{PRC103044315}   &       	&13.680 	&				\\	
14.1 (1) 	&220 (20) 	&0.06  	&0.94 	&   	&		&				&   	&       	&14.084 	&not 2$^{+}$	\\
14.3 (1) 	&90 (20) 	&0.47 	&0.53 	&   	&		&				&		&			&14.348 	&3-			\\	
14.8 (1) 	&150 (30) 	&0.02 	&0.98 	&   	&		&				&		&			&14.863 	&2$^{+}$		\\  	
15.3 (1) 	&240 (20) 	&0.13 	&0.87 	&   	&15.3 	&0$_{8}^{+}$	&15.33 (3)~\cite{PRC103044315}	&			&15.347 	&$4^{+}$		\\	
16.7 (1) 	&340 (10) 	&0.11 	&0.83 	&0.06 	&		&				&		&			&16.666 	&even			\\
17.2 (1) 	&300 (10) 	&0.19 	&   	&0.81 	&		&				&		&			&17.133 	&5$^{-}$		\\	
17.6 (1) 	&290 (80) 	&1.00 	&   	&   	&		&				&		&			&17.615 	&5$^{-}$		\\	
17.9 (1) 	&90 (70) 	&0.26 	&0.58 	&0.16 	&		&				&		&			&17.830 	&not 4$^{+}$	\\
18.1 (1) 	&120 (90) 	&0.28 	&0.64 	&0.08 	&		&				&		&			&18.149 	&5$^{-}$		\\
18.3 (1) 	&140 (120) 	&0.15  	&0.85 	&   	&		&				&		&			&18.320 	&0$^{+}$,6$^{+}$\\
19.3 (1) 	&300 (20) 	&   	&0.30 	&0.70  	&		&				&		&			&           &				\\  		
20.1 (1) 	&300 (50) 	&0.14 	&0.48 	&0.38 	&		&				&		&			&           &				\\  		
20.3 (1) 	&180 (10) 	&0.06 	&0.03 	&0.91 	&		&				&		&			&           &				\\  		
20.6 (1) 	&100 (40) 	&0.13	&0.13 	&0.74 	&		&				&      	&			&			&				\\  		
21.0 (1) 	&510 (60) 	&   	&   	&1.00 	&		&				&		&			&           &				\\  		
21.5 (1) 	&380 (20) 	&0.08 	&0.61 	&0.31 	&		&				&      	&			&			&				\\  		
21.7 (1) 	&140 (60) 	&0.12 	&0.44	&0.44 	&		&				&		&			&           &				\\  		
21.9 (1) 	&370 (20) 	&0.09 	&0.11 	&0.80 	&   	&           	& 	    &			&			&				\\  		
22.9 (1) 	&480 (50) 	&0.63 	&0.37 	&   	&		&				&		&			&           &				\\  		
23.1 (1) 	&190 (50) 	&0.03 	&0.88 	&0.09 	&		&				&		&			&           &				\\  		
23.4 (1) 	&320 (40) 	&   	&0.55 	&0.45 	&		&				&		&			&           &				\\  		
24.4 (1) 	&890 (80) 	&0.06 	&0.55 	&0.39 	&		&				& 	    &			&			&				\\  		
24.5 (1) 	&480 (30) 	&0.10 	&    	&0.90 	&   	&           	&		&			&	        &				\\  		
25.2 (1) 	&380 (36) 	&0.12 	&0.52 	&0.36	&    	&           	&		&			&           &				\\  		
26.1 (1) 	&630 (40) 	&0.02 	&0.29   &0.69 	&		&				&		&			&           &				\\  		 		
\bottomrule
\end{tabular*}%
\end{table*}

Two forward telescopes (L0 and R0) were centered at 24$^{\circ}$ with respect to the beam direction and placed at a distance of 162.25 mm from the target center. Each consisted of two layers of double-sided silicon strip detectors (DSSDs) with thicknesses of approximately 40 $\mu$m (W1) and 500 $\mu$m (BB7) and one large-size silicon detector (SSD) with a thickness of approximately 1500 $\mu$m (MSX40). The active areas were 50 mm $\times$ 50 mm for W1 and 64 mm $\times$ 64 mm for BB7 and MSX40. For the DSSDs, the front and back sides were segmented into 16 (for W1) or 32 (for BB7) strips, providing good two-dimensional position resolutions and the ability to record multi-hit events in one telescope. L1/R1 had the same composition as L0/R0 but were centered at 47$^{\circ}$ and placed at a distance of 176.25 mm. L2 (or R2) was composed of one DSSD-W1 and one SSD and centered at 78$^{\circ}$. L3 (or R3)  consisted of one DSSD-BB7 and one SSD, centered at 110$^{\circ}$.

The energy match for all silicon strips in one DSSD can be achieved using the uniform calibration method described in Ref.~\cite{IEEE61596}. Absolute energy calibration was performed using an $\alpha$-particle source with three energy components (5.155, 5.486, and 5.805 MeV) and an elastically scattered $^{7}$Li particle from a thin $^{197}$Au target~\cite{PRC99064315}. The typical energy resolutions of the silicon detectors were less than 1.0\% for a 5.805-MeV $\alpha$ particle. The position resolutions can be determined by the respective strip widths.

The timing signals from the strip detectors were also recorded, which can be used to discard most accidentally coincident events. This is important because the strips close to the beam axis sustained a high counting rate. As depicted in Fig.~\ref{fig1}(b), good particle identification performance has been achieved for light isotopes after applying suitable conditions on the coincidence time, as well as those on the track information across several silicon layers~\cite{PRL124192501}.

\begin{figure}[!htb]
\includegraphics[width=\hsize]{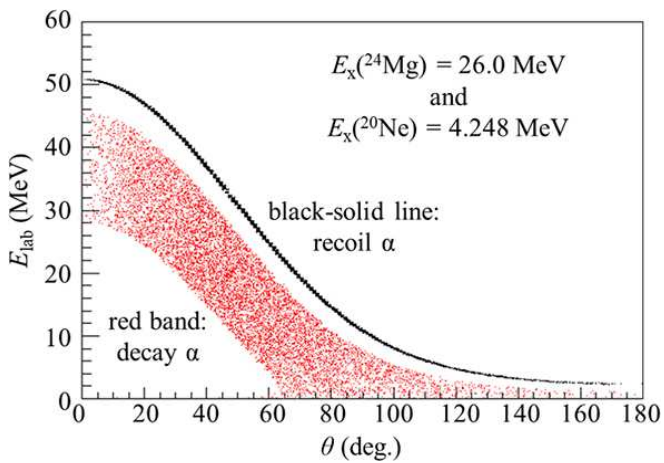}
\caption{(Color online) Relationship between energy and angle (Lab) for the recoil $\alpha$ particles (black solid line) and decay $\alpha$ particles originating from excited $^{24}$Mg (26 MeV) and accompanied by $^{20}$Ne (4.248 MeV) (red band).}
\label{fig2}
\end{figure}

Two $\alpha$ particles from the targeted reaction $^{12}$C($^{16}$O,$\alpha$+$^{20}$Ne)$\alpha$, namely, the decay and recoil particles, can also be discriminated to a large extend according to kinematics, subject to a limit on the excitation energy. Figure~\ref{fig2} shows the simulated energy vs angle distribution for the two types of $\alpha$ particles, assuming an excitation energy of 26.0 MeV in $^{24}$Mg and 4.248 MeV in the $^{20}$Ne fragment. In the plot, the black solid line represents the recoil $\alpha$ particles, whereas the red band represents the decay $\alpha$ particles. The black line tended to approach the red band when the excitation energy in $^{24}$Mg was increased. The critical energy appeared at approximately 30 MeV, below which the two sources of the $\alpha$ particles were clearly distinguished. Hence, in the present study, we only extracted resonances below 30 MeV.

Once two out of three final-state particles were identified using the $\Delta E-E$ method, the kinematic quantities of the third particle could be deduced according to the energy and momentum conservation~\cite{PRC64044604}. For the reaction $^{12}$C($^{16}$O,$^{24}$Mg$\rightarrow$$\alpha$+$^{20}$Ne)$\alpha$, the heavier $^{20}$Ne fragment could not be directly identified from the detection because it should have been stopped in the first DSSD layer. Therefore, the resonant states of $^{24}$Mg could only be reconstructed using the detected decay $\alpha$ particle and deduced $^{20}$Ne fragment.

\section{ANALYSIS AND RESULTS}\label{sec.III}

\begin{figure}[!htb]
\includegraphics[width=\hsize]{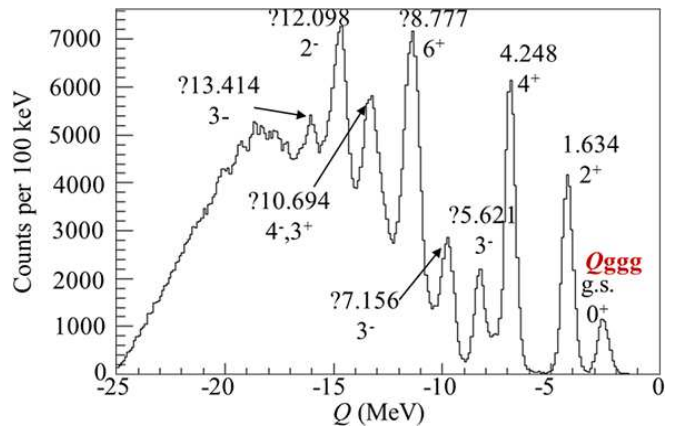}
\caption{(Color online) $Q$-value spectrum for the reaction $^{12}$C($^{16}$O,$^{24}$Mg$\rightarrow$$\alpha$+$^{20}$Ne)$\alpha$, calculated using the energies of the detected decay and recoil $\alpha$ particles and the deduced $^{20}$Ne fragment. The peaks in the spectrum are associated with the ground and excited states (as marked) of the $^{20}$Ne fragment.}
\label{fig3}
\end{figure}

The reaction $Q$ value is defined by the mass deficit between the initial and final particles and is therefore useful in determining the reaction channels. Equivalently, it can be calculated from the energy released during the reaction:
\begin{eqnarray}
\label{eq1}
Q = E_{{\rm recoil}\_\alpha}+E_{{\rm decay}\_\alpha}+E_{^{20}{\rm Ne}}-E_{^{16}{\rm O}}.
\end{eqnarray}
where $E_{^{20}{\rm Ne}}$ represents the deduced value. Figure~\ref{fig3} shows the excellent resolution of the $Q$-value spectrum in which the ground state ($Q_{{\rm ggg}}\sim$ 2.540 MeV), first excited state (excitation energy at $E_{x}\sim$ 1.634 MeV and spin-parity of 2$^{+}$), and second excited state ($E_{x}\sim$ 4.248 MeV, 4$^{+}$) of $^{20}$Ne can be clearly discriminated, corresponding to different decay paths. For $E_{x}$($^{20}$Ne) $ > $ 5 Mev ($Q < -7$ MeV), there are many close-by states in $^{20}$Ne, which were indistinguishable in the $Q$-value spectrum and labeled by "?" in the figure. We note that the good $Q$-value resolution is attributed to the excellent energy resolutions for the incident beam and the silicon detector, as well as to the small energy loss in the target~\cite{PRC95021303}.

Using the detected decay $\alpha$ particle and the deduced $^{20}$Ne fragment, the relative energy (or decay energy) of the resonances in $^{24}$Mg can be reconstructed according to the standard invariant mass (IM) method~\cite{PRL112162501,PRC95021303,CPC42074003,PRL124192501}. A contamination channel arises from the reaction $^{12}$C($^{16}$O,$^{8}$Be$\rightarrow$$\alpha$+$\alpha$)$^{20}$Ne, which has the same final mass combination and hence cannot be eliminated by the $Q$-value selection. We checked this contamination channel using the two-dimensional Dalitz-plot~\cite{SCPM6212011,PM441068} corresponding to the reconstruction of $^{24}$Mg versus $^{8}$Be. It was found that the formation of $^8$Be presents some background for $^{24}$Mg-states at $E^* >$ 20 MeV. In the data analysis, we cut off events with $E_{^{8}\rm{Be}}^* <$ 3 MeV to improve the signal-to-background ratio. The same cut was also applied to the efficiency simulation (see below).

In Fig.~\ref{fig4}(a-c), we plotted three excitation-energy (relative energy plus the corresponding separation energy) spectra for $^{24}$Mg , conditioned by the highest three $Q$-value peaks, as shown in Fig.~\ref{fig2}. Each spectrum was fitted by a number of resonance peaks plus a smooth varying continuum background ~\cite{PRC94034313, PRC95021303}. The peak positions were initialized according to the previously reported results, the actually obtained spectrum shape, and the consistency between spectra for different decay paths, as presented in Fig.~\ref{fig4}(a), (b) and (c). During the fitting procedure, the corresponding peak centroid and width for one resonance in $^{24}$Mg decaying into different final states of $^{20}$Ne were kept the same. Each  peak in the figure is a convolution of the Breit-Wigner (BW) form with the Gaussian-type energy-resolution function~\cite{PRL124192501}. The energy resolution and the detection efficiency (acceptance) curves as a function of the relative energy were obtained via Monte Carlo simulation considering a reasonable angular distribution of the produced $^{24}$Mg and its decay fragments as well as the actual detection setup, energy and position resolutions of the detectors, and applied cuts in the data analysis~\cite{PRC95021303,PRL124192501}. Because the efficiency curves (Fig.~\ref{fig4}) varied gradually over the range of a typical peak, we maintained it as a constant for each peak~\cite{PRL124192501}. The extracted resonance energies ($E_x$) and widths ($\Gamma$) are listed in Table \ref{tab1} in the first and second columns, respectively. The values in the parentheses give the corresponding statistical errors (standard deviation). In addition, a systematic uncertainty was estimated according to the Monte Carlo simulation, which was approximately 100 keV for both the peak centroid and width~\cite{PRC95021303}. Note that the statistical significance for each peak in Fig.~\ref{fig4} was also evaluated by taking the ratio of the peak count to the square root of the background count (statistical fluctuation $\sigma$) within a range of $\pm \Gamma$ relative to the peak centroid. In the case of one resonance decaying into various final states of $^{20}$Ne, only the most significant was taken into account. In Table~\ref{tab1} and Fig.~\ref{fig5}, we present only the resonant states with those significantly larger than 5$\sigma$ (confidence level (CL) $> 99.99\%$).

\begin{figure}[!htb]
\includegraphics[width=\hsize]{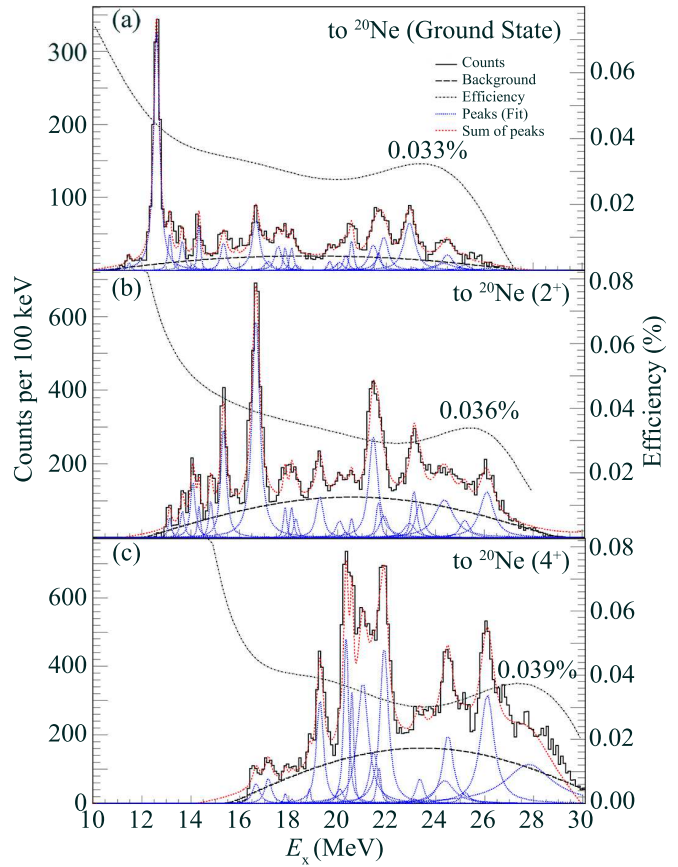}
\caption{(Color online) Excitation-energy (relative energy plus the corresponding separation energy) spectra conditioned by the highest three $Q$-value peaks, as indicated in Fig.~\ref{fig3}. Each spectrum is fitted by a number of peak functions (blue-dash lines), which are BW forms convoluted with energy resolution functions, plus a continuum background (black-dashed lines). The simulated detection efficiency curves (black-dotted lines) are also plotted, with several characteristic values indicated at the curves.}
\label{fig4}
\end{figure}

From the reconstructed IM spectra in Fig.~\ref{fig4}, the relative decay width for each resonant state to one specific final state, as a ratio to the sum of the widths for all three final states, can be extracted according to the number of counts in each fitted peak, corrected by the corresponding detection efficiency. The results are shown in Fig.~\ref{fig5} and listed in Table \ref{tab1}.

\begin{figure*}[!htb]
\includegraphics[width=16 cm]{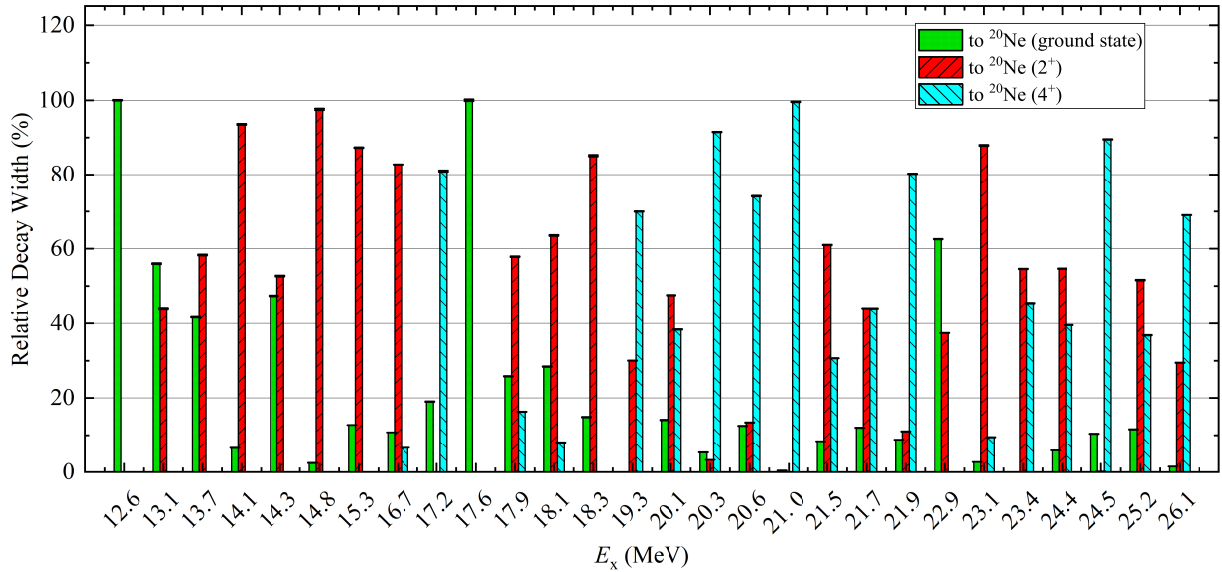}
\caption{(Color online) Relative decay width determined from the numbers of counts in each resonance peak, corrected by the corresponding detection efficiency, and normalized to the sum of the widths for all three final states of $^{20}$Ne. The green-filled, red-upward-hashed, and blue-downward-hashed bars represent the relative widths for decaying into the ground, 1.634 MeV (2$^{+}$), and 4.248 MeV (4$^{+}$) states of $^{20}$Ne, respectively.}
\label{fig5}
\end{figure*}

\section{DISCUSSION} \label{sec.IV}

The observed resonances for $E_{x} \textless 18.5$ MeV effectively reproduced the results of a previous $\alpha$-cluster resonant elastic scattering experiment (see Ref.~\cite{PRC432523} and Table \ref{tab1}), confirming the correctness of our measurement and data analysis. In the $^{24}$Mg($\alpha$,$\alpha ^{'}$) inelastic scattering experiments by Kawabata $\it et\ al$ ~\cite{JPCS436012009} and Adsley $\it et\ al$ ~\cite{PRC103044315,PRL129102701}, the three states at $E_{x}$ = 13.1, 13.79 and 15.33 MeV in $^{24}$Mg were found to have large $\alpha$-decay widths compared to their corresponding proton-decay widths and were tentatively assigned spin-parities of $0^+$. These states have correspondence to our measurement and are presented in Table \ref{tab1}. On the theoretical side, the work of Chiba and Kimura using the antisymmetrized molecular dynamics (AMD) approach combined with the generator coordinate method (GCM) has predicted a series of $0^+$ states with various cluster configurations for $^{24}$Mg~\cite{PRC91061302}, which are also listed in Table~\ref{tab1} for comparison. The present study showed that the $\alpha$+$^{20}$Ne configuration in $^{24}$Mg may persist to high excitation energies, considerably beyond the thresholds of other cluster configurations. This may affect the formation of other types of clusterings, which should be orthogonal to existing states.

Recently, decay-path selection was proposed and successfully applied to probe the specific cluster structure ~\cite{PRL124192501, PRC105044302}. In the present study, owing to the excellent resolution on $Q$-value spectrum and thus the clear distinction of the decay paths for each event, we were able to extract the relative decay strengths of each $^{24}$Mg resonance, as displayed in Fig.\ref{fig5} and listed in Table~\ref{tab1}. Interestingly, we found many resonances decaying dominantly into the first or second excited states of the $^{20}$Ne fragment instead of into its ground state. This may provide a test ground for future theoretical calculations.

\section{Summary} \label{sec.V}

A transfer reaction and cluster decay experiment, $^{16}$O($^{12}$C,$^{24}$Mg$\rightarrow$$\alpha$+$^{20}$Ne)$\alpha$, was conducted at a beam energy of 96 MeV. This reaction channel exhibited a large $Q$-value in favor of populating the highly excited states of $^{24}$Mg. The obtained $Q$-value spectrum exhibited a sufficiently high resolution, allowing us to discriminate the decay paths collected to the ground, first excited (1.634 MeV), and second excited (4.248 MeV) states of the $^{20}$Ne final fragment. A number of resonances in $^{24}$Mg with the $\alpha$+$^{20}$Ne cluster configuration were observed up to an excitation of 30 MeV, together with the total decay width and relative decay strengths for various decay paths. These experimental results encourage more theoretical work to correctly reproduce the $\alpha$+$^{20}$Ne structure in $^{24}$Mg before being applied to other more complicated configurations.

\acknowledgments{The authors thank the staff of the HI-13 tandem accelerator at the CIAE for providing excellent technical and operational support during the experiment.}

\end{document}